\documentclass[twocolumn,showpacs,preprintnumbers,amsmath,amssymb]{revtex4}
\usepackage{epsfig}
\def\ben{\begin{enumerate}}
\def\een{\end{enumerate}}
\def\bit{\begin{itemize}}
\def\eit{\end{itemize}}
\def\beq{\begin{equation}}
\def\eeq{\end{equation}}
\def\bea{\begin{eqnarray}}
\def\eea{\end{eqnarray}}
\def\bq{\begin{quote}}
\def\eq{\end{quote}}
\def \lsim{\mathrel{\vcenter
     {\hbox{$<$}\nointerlineskip\hbox{$\sim$}}}}
\def \gsim{\mathrel{\vcenter
     {\hbox{$>$}\nointerlineskip\hbox{$\sim$}}}}
\def\gappeq{\mathrel{\rlap {\raise.5ex\hbox{$>$}}
{\lower.5ex\hbox{$\sim$}}}}
\def\lappeq{\mathrel{\rlap{\raise.5ex\hbox{$<$}}
{\lower.5ex\hbox{$\sim$}}}}

\def\hats{\hat{s}}
\def\hatt{\hat{t}}

\def\epem{e^+e^-}
\def\mpmm{\mu^+\mu^-}
\def\lplm{\ell^+\ell^-}

\def\m3e{\mu \to e \bar{e} e}
\def\a{\alpha}

\def\g{\gamma}

\def\m{\mu}

\begin{document}

%\preprint{APS/123-QED}

\title{Of Contact Interactions and Colliders }% Force line breaks with \\

\author{ Sacha Davidson}
 \email{s.davidson@ipnl.in2p3.fr}
\affiliation{ 
IPNL, Universit\'e de Lyon, Universit\'e Lyon 1, CNRS/IN2P3, 
4 rue E. Fermi 69622 Villeurbanne Cedex, France}
\author{S\'ebastien Descotes-Genon}
 \email{sebastien.descotes-genon@th.u-psud.fr}
\affiliation{Laboratoire de Physique Th\'eorique, CNRS/Univ. Paris-Sud (UMR 8627), 91405 Orsay Cedex, France}
\author{Patrice Verdier}
 \email{verdier@ipnl.in2p3.fr}
\affiliation{ 
IPNL, Universit\'e de Lyon, Universit\'e Lyon 1, CNRS/IN2P3, 
4 rue E. Fermi 69622 Villeurbanne cedex, France}
%\affiliation{ }

%\date{\today}% 

\begin{abstract}
The  hierarchy  of scales 
which would allow dimension{-}six contact interactions 
to parametrise New Physics   
may not be verified at colliders. Instead, we explore the feasability 
and usefulness of  parametrising the high{-}energy tail of
distributions at the LHC using  form factors.
We focus on the process 
$pp \to \ell\bar{\ell}$ in
the presence of  $t$ (or $s$)-channel New Physics, 
guess a form factor from the  partonic cross-section, and
attempt to use data to constrain its coefficients, and  
 the coefficients to  constrain models. We find that our
choice of form factor decribes $t$-channel exchange
better than a contact interaction, and the coefficients
in a particular model can be obtained from the partonic cross-section.
We estimate bounds on the coefficients by fitting
the form factors to available data.  For the
parametrisation corresponding to the contact interaction
approximation,  our expected  bounds on the 
scale $\Lambda$ are  within $\sim 15\%$ of the 
latest limits from the LHC experiments.
\end{abstract}

\maketitle

\section{Introduction}

Suppose that the LHC does not discover  new particles in
direct production. It can
nonetheless be sensitive to  new particles
just beyond its kinematic reach,  from their effects
on the high energy  tails of  distributions. 
These effects are usually parametrised by contact
interactions, which are local, non-renormalisable
operators. For instance, the
process $pp\to \lplm$ can be
sensitive to the four fermion  operator
\beq
\label{ATLAS}
\pm \frac{4\pi}{\Lambda^2}\sum_{q=u,d}(\overline{q}\g^\a P_L q ) (\overline{\ell}\g^\a P_L \ell)
\eeq
where $\ell=e,\mu$. 
 With 20 fb$^{-1}$ of data at  
8 TeV in the centre-of-mass frame, LHC experiments
have set bounds on some  coefficients $4\pi/\Lambda^2$
%of four-fermion contact interactions 
of order
$\Lambda \gsim  10-17$ TeV \cite{ATLAS,CMSLam, CMSqang,CMSqs}.
Various recent papers \cite{CIth}
have explored  what can be learned from contact
interaction studies, if the LHC finds no new particles.

Two difficulties arise in
attempting  to apply currently available contact interaction (CI)
 bounds
to specific New Physics (NP) Models:\\ 
{\bf 1)}
Experimental  limits exist only for a selection of CIs,
among the large collection
 labelled by the chirality,
flavour and gauge charge  of participating fermions, as
well as the Lorentz structure of the interaction.  Since the
magnitude and sign of the
interference with the Standard Model (SM)  depends on these labels,
it is  improbable that an available limit will be applicable
to the   interactions induced by
a  particular  model.
\\
{\bf 2)} In the sensitivity range of colliders,
it is unlikely that the CI
approximation (that $p^2 \ll m^2$ for the heavy mediating
particle) is satisfied in any but the most strongly coupled
of models.

A  non-local, or  ``form factor''  parametrisation
of the distribution tails might address
both points:  with
a judicious choice of  functional form,
it  may include non-local interactions
mediated by propagating particles, and 
its coefficients may be  simply calculated in many models. 
In this paper, we focus on the process $q\bar{q}\to \ell^+
\ell^-$ at the LHC, and 
parametrise the $pp \to \lplm$ cross-section 
as:
\beq
\frac{d\sigma }{d\hats} = 
\frac{d \sigma_{DY} }{d\hats} 
\left(1  +
a\frac{\hats }{1+ c\hats} + 
b\frac{\hats^2}{(1 + c\hats)^2}
\right)  ~~~,
\label{dream}
\eeq
where  $a$, $b$, and  $c$
are coefficients to be determined, respectively of mass dimension 
-2, -4, -2, 
%and while $a$ can be negative, $b> 0$. where 
$\sigma_{DY}$ is the Drell-Yann (DY) cross-section
for $Z/\g$ exchange\footnote{The Drell-Yann cross-section depends on 
parton distribution functions (pdfs),
so bounds on NP extracted this way  suppose
that the pdfs are reliably obtained in some
other process. From a theoretical 
perspective, the pdf dependence could
be avoided by studying $\frac{d\sigma }{d\hats}(\to \epem)/
\frac{d\sigma }{d\hats}(\to \mpmm)$. This would
be sensitive to New Physics which
coupled differently to $e$s and $\mu$s.},  and $\hats$ is the invariant
mass-squared of the final state leptons.

Section \ref{sec:CIvsFF} supports the functional
form of  eq. (\ref{dream}) by 
 studying the partonic cross-section for
$t$-channel exchange of  a leptoquark
with mass just beyond the reach of the LHC.
Then  section  \ref{sec:ptoFF}  argues
that for  a generic
model,  $a,b$ and $c$ can be estimated 
from the partonic cross-section with
simple approximations to the
parton distribution functions (pdfs). 
Finally,  section
\ref{sec:data?}  attempts a least-squares
fit of eq.
(\ref{dream})  to available data.

Constraints on leptoquarks \cite{LQ,WZ,assia}
and contact interactions involving two quarks and
two leptons have been widely studied \cite{tlmonde},
 both from precision and collider data. It is
generically true that colliders  have the best
sensitivity to  flavour-diagonal  operators. A
parametrisation of contact interactions similar to
eq. (\ref{dream})  with $c\to 0$,  was 
proposed in \cite{Jose} to address the first
problem above.  Our form-factor generalisation is
perhaps an old-fashioned version of  ``simplified models''\cite{SimMod}.

\section{Form Factors vs Contact Interactions}
\label{sec:CIvsFF}

The aim of this section is to justify replacing $4\pi/\Lambda^2$
as the coefficient of a four-fermion operator, by the
non-local coefficient $ \sim \lambda^2/(m^2+\hats)$. 
We focus on a particular model, a
 scalar   leptoquark  $S$  with
interaction $\lambda S \overline{e}P_L u^c + h.c. $\cite{LQ,assia},
and argue that the partonic cross-section
is better approximated by  an
%In this section, we consider the $t$-channel exchange of 
%a heavy leptoquark of mass $m$, and  find  that
%the form factor 
expansion in $\hats/(\hats +m^2)$
% converges better 
than 
%the CI expansion 
in  $\hats/m^2$. We imagine this
could be generalised to any particle  exchanged in 
the $t$-channel, and will discuss multiple
particle exchange in a later publication.

The   partonic cross-section
 for $u\bar{u} \to \epem$, mediated by 
 neutral gauge bosons  and $t$-channel  $S$ exchange, is:
\bea
\! \! \! \! \! \! \!
\hat{\sigma}_{LQ}\! \! \! &=& \! \! \!
\hat{\sigma}_{DY} +
\frac{1}{48 \pi \hats}
\left[
-\frac{2 g^{'2}\lambda^2}{3}
\left(\frac{1}{2} -\frac{m^2}{ \hats} 
  +\frac{m^4}{\hats^2} \ln (\frac{m^2+\hats}{m^2}) \right) \right.
\nonumber \\
&&\left. +
\frac{\lambda^4}{4}
\left(\! \!1 -2\frac{m^2}{\hats} \ln (\frac{m^2+\hats}{m^2})   
+\frac{m^2}{(m^2 + \hats)}  \right)
\right]
\label{sighLQ}
\eea
where the gauge bosons are taken 
to be the massless $W_0,B$  with couplings
$g_2,g'$, because  $\hats \gg m_Z^2$.
%\beq \hat{\sigma}_{DY} =... \label{hatDY} \eeq

\begin{figure}[ht]
\unitlength.5mm
%\SetScale{1.418}
%\begin{boldmath}
\begin{center}
\epsfig{file=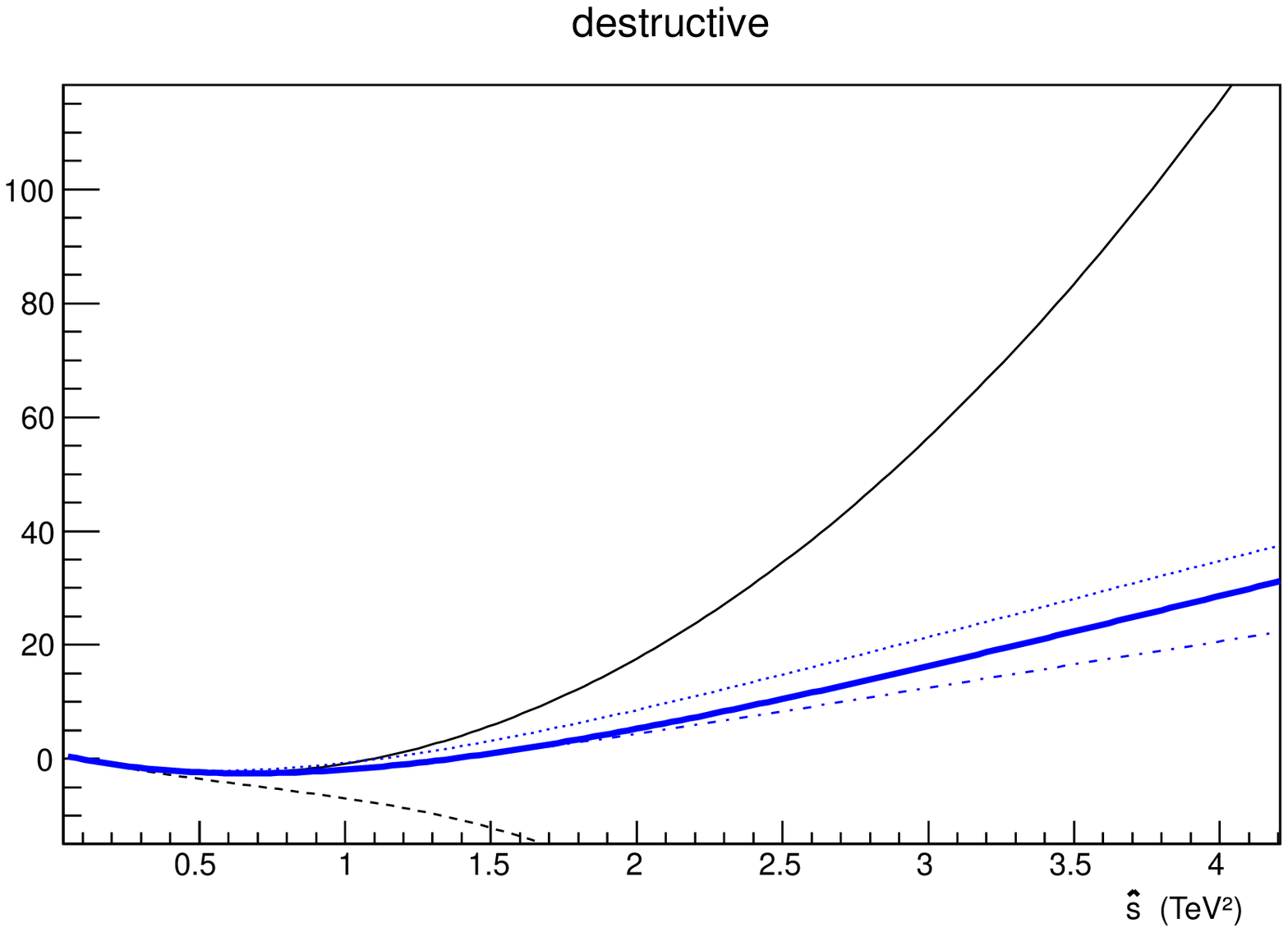,height=4cm,width=8cm}
\epsfig{file=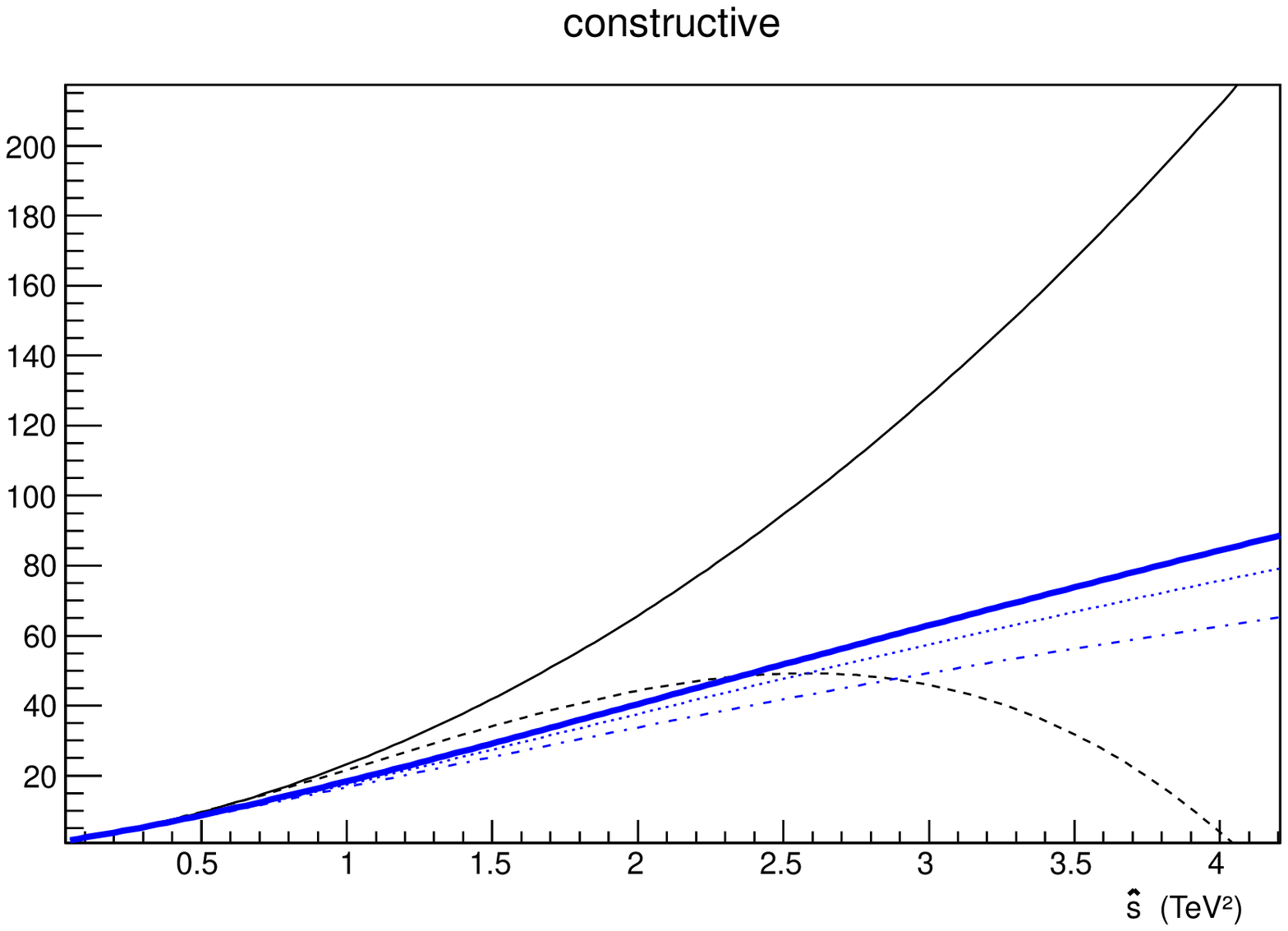,height=4cm,width=8cm}
\end{center}
%\end{boldmath}
\caption{ 
Thick blue is the partonic cross-section
for $\hat{\sigma}(u\bar{u} \to \epem)$, mediated by
$Z,\gamma$ and a leptoquark $S_0$
of mass 2 TeV and  $\lambda^2 = 1$
 in the  $t$-channel, and normalised
to  $\hat{\sigma}$ for $Z,\gamma$ exchange.
The dotted (dot-dashed) blue   are
the  ${\cal O} (\hats^2), 
{\cal O} (\hats^3)$ terms of a ``form factor'' approximation
 to leptoquark exchange (see eq. (\ref{LQff})).
Black solid (exits the upper
plot border)  is the  
dimension-six CI  approximation,
and black dashed (exits lower border) 
 includes the first correction in $\hats/m^2$ to
the CI approximation. The
leptoquark has  destructive
interference (first plot), for illustration,
we  also plot the same model  but with  constructive
interference.  
% Fait avec ORDI/ROOT/LQ/rxfin.C  
\label{fig:rxs} }
\end{figure}

The ratio $ \hat{\sigma}_{LQ}/\hat{\sigma}_{DY}$
is the solid blue line in figure \ref{fig:rxs}.
To obtain the form factor parametrisation, we expand 
\beq
\ln  \left(\frac{m^2+\hats}{m^2}\right) 
%=  -\ln \left( 1 -\frac{\hats}{m^2+\hats}\right)
\simeq \frac{\hats}{m^2+\hats} + \frac{\hats^2}{2(m^2+\hats)^2}
%+ \frac{\hats^3}{3(m^2+\hats)^3} 
+ ...
\eeq
The first  terms in  
expanding   $144\pi \hats (\hat{\sigma}_{LQ} - \hat{\sigma}_{DY})$   are:
\bea
%\! \! \! &=& \! \! \!
%\hat{\sigma}_{DY} 
%\frac{1}{144 \pi\hats }\left[ 
 \frac{-2 g^{'2} \lambda^2 }{3}
\frac{\hats}{m^2 + \hats}  
+\left(
 \frac{ -g^{'2} \lambda^2 }{6}
+ \frac{ \lambda^4 }{4}
\right)\frac{\hats^2}{(m^2 + \hats)^2}
+...
%\right]~
\label{LQff}
\eea
and are plotted in figure \ref{fig:rxs}.  
Already two terms give a reasonable
fit to the exact result, and are  
only slightly more
complicated than a CI, so we opt for 
the simple form of  eq. (\ref{dream}),
despite that
in principle, form factors can be
the most general functions of the available variables,
restricted  only by  symmetries. 

Notice that for $\hats \ll m^2$, we could also
expand $\ln (1 + \hats/m^2) \simeq \hats/m^2 - \hats^2/2m^4 + ...$,
which corresponds to parametrising leptoquark
exchange by a tower of  local operators, 
 including dimension-eight CIs $\propto \hats/m^4$.
But this expansion is less useful,  because
it fails  in the regime  $\hats \gsim m^2$. 
This is illustrated in
figure \ref{fig:rxs} by the black lines exiting the
top and bottom of the plots. The continous black line is
$\hat{\sigma}_{LQ}/\hat{\sigma_{DY}}$ expanded to 
second order in $\hats/m^2$ (the CI approximation),
and the dashed line includes
third order in
$\hats/m^2$.
This poor convergence motivates our interest in form factors,
as opposed to CIs. So we are
abandonning the theoretical attractions of
the Operator Product Expansion and local Effective
Field Theory\cite{Georgi}, because there is insufficient hierarchy
between $\hats$ and $m^2$ to justify
truncating the  expansion in local operators
at the lowest orders.

A curious feature of eq. (\ref{LQff}) is the
 ${\cal O} (g^2 \lambda^2)$
 contribution to
the   ${\cal O} (\hats^2)$ term,
which  is usually \cite{CCS} absent  in the  
CI approximation.
%It  arises  from
%interference of  an ${\cal O}(1/m^4)$ 
%NP interaction with the Drell-Yan amplitude,
%and should be a small correction
This neglect is justified,   because the CI limit 
is $m^2 \gg \hats$, and  colliders  are sensitive
to CI with $\lambda^2/2m^2 \gsim g^2/\hats$,
which implies $g^2 \ll \lambda^2$.
In the case of form factors,  it is convenient to  neglect
this contribution in our preliminary
analytic estimates eqns (\ref{8b}) -(\ref{eq2.1.14}),
but  we will see
that it should be included.

We now briefly comment on the $s$-channel exchange of 
 a new particle (or resonance) of mass $M$, which  generates
a peak in
the partonic cross-section  at $\hats = M^2$.
The rise towards this peak, for $\hats \ll M^2$ 
can be parametrised as a contact interaction
(eq. (\ref{dream}) with $c\to 0$). However,
the expansion in $\hats/(\hats + M^2)$, which
was useful for $t$-channel exchange, has
no advantages in this case. For  $M^2 <\hats$,
it is possible that 
an $s$-channel resonance  could contribute
a shoulder (like  $t$-channel exchange)
in the binned $pp\to \lplm$ data. However,
this depends on pdfs, binning and the particle's
properties, so we  will discuss 
 in a later publication the prospects
of extracting the properties of the resonance
from  a form factor parametrisation of the cross-section.

\section{From partons to form factors}
\label{sec:ptoFF}

This section aims to relate the
partonic cross-section  $\hat{\sigma}(\bar{q}q \to \lplm)$,
mediated by  arbitrary New Physics, 
to the
differential cross-section for $pp \to \lplm$:
\bea
\frac{d\sigma}{d\hats} = 
\frac{2}{s}
\sum_{q=u,d}
 \int   d\eta^+
\, d\hatt 
 f_q(x_1)  f_{\bar{q}}(x_2) 
\frac{d \hat{\sigma}}{d\hatt}(\bar{q}q \to \lplm)
\label{sigma}
\eea
where $f_q$ is  the pdf   of
the quark $q$, and
$x_1 = \frac{M_{\epem}}{\sqrt{s}} e^{\eta_+}$,  %~~~,~~~ 
$ x_2 = \frac{M_{\epem}}{\sqrt{s}} e^{-\eta_+}$ 
 are the fractions  of the proton's momentum carried 
by the colliding partons. %, and  there is a 2 because the
%valence quark could be in either incident proton. 

Our first approximation is to suppose a single
density for sea quarks, and  that there are twice
as many valence $u$s as $d$s.
Concretely, we take:
\bea
\!\!\!   \!\! \int\!\! d\eta_+ f_u(x_1)f_{\bar{u}}(x_2)\!\!\!  &=&
%\nonumber\\
\!\!\!   \frac{2}{3}  \sum_{q=u,d} \!\! \int d\eta_+ f_q(x_1)f_{\bar{q}}(x_2) 
%+ f_d(x_1)f_{\bar{d}}(x_2))  
\equiv \!\!   \frac{2}{3} F(\hats)
\nonumber\\
\!\!\!  \int \!\! d\eta_+ f_d(x_1)f_{\bar{d}}(x_2) \!\!\!   & =&
%\nonumber\\
\!\!\!   \frac{1}{3} \sum_{q=u,d}\!\!\int d\eta_+ f_q(x_1)f_{\bar{q}}(x_2)~. 
%+ f_d(x_1)f_{\bar{d}}(x_2))  \equiv  \frac{1}{3} F(\hats)
\label{poid}
\eea
The  second approximation is to take 
simple  integration limits $[0 , \hats]$  for  
%$\eta_+ =(\eta_{e} + \eta_{\bar{e}})/2 $ and 
$-\hatt$ , despite
that experiments restrict the
angular distribution to be
away from the beam-pipe. This should
be acceptable if the NP and SM events
have the same angular distribution, as is the
case for the $V\pm A$ four-fermion interactions
\footnote{Non-$V-A$ operators will be discussed in
a subsequent publication.  Their cross-section
 may have
a different angular  distribution, which for instance  could affect
event migration among bins. However, since the difference
 is only in the numerator (upstairs), it should not induce
dangerous effects like a divergence along the beampipe.}.

These approximations  allow to factorise
the  $pp\to \lplm$ cross-section as a pdf integral multiplying
an ``averaged'' partonic cross-section
$\widetilde{\sigma} =
\frac{2}{3}\hat{\sigma} (\bar{u} u \to \lplm)
+\frac{1}{3}\hat{\sigma} (\bar{d} d \to \lplm)$:
\bea
\frac{d \sigma}{d \hats}  &\simeq& 
\frac{2 F(\hats)}{s} \widetilde{\sigma}%\nonumber\\
\label{LQCInotn} 
\eea
where, for  Drell-Yann, with   $\sin^2\theta_W= 1/4$:
\bea
\widetilde{\sigma}_{DY}
  &\simeq& 
 \frac{g_2^4 }{1280 \pi \hats }~~.
\eea
The expected SM rate includes non-DY processes (dibosons, etc)
which amount to 10-20\% of the rate in the data to which  we
will compare\cite{CMSLam}. 
We augment $\widetilde{\sigma}_{DY}$ by $\sim 10\%$
to account for this, which
 allows to write  $\frac{d \sigma}{d \hats}$ in the presence of NP as
\bea
\frac{ F(\hats)}{72\pi s}  \!\! \left( 
 \frac{ g_2^4 }{8  \hats }\!
\! + \! \epsilon_{int}g_2^2  \frac{4\pi/\Lambda^2}{(1 +\hats/m^2) } 
\!\!+ \!
\epsilon_{NP} 
\frac{16\pi^2/\Lambda^4}{ (1 +\hats/m^2)^2 }  {\hats} \! \right)
\label{8b}%\\
\eea
where $4\pi/\Lambda^2$ %$ \lambda^2/2m^2$
  is the coefficient of the
CI induced by the NP,  and 
$\epsilon_{int}$ and $\epsilon_{NP}$ are constants,
predicted by the model and obtainable from
the partonic cross-section, which respectively parametrise
the SM-NP intereference, and  account for the
partons involved in the NP interactions. 
 The $\epsilon_{int}$ and $\epsilon_{NP}$ for  leptoquarks
 are tabulated in \cite{assia};
the leptoquark  considered  here (see eq. \ref{LQff}) has:
\bea 
\epsilon_{NP}=2/3 ~~,~~\epsilon_{int} =-8/27~~~,~~\frac{4\pi }{ \Lambda^2} = \frac{ \lambda^2 }{2m_{LQ}^2}~~~, 
\eea
and the interaction (\ref{ATLAS}) has
$\epsilon_{NP}=1$, $\epsilon_{int} =\mp 1/6$.
These parameters can be translated   to the $a,b,c$ 
which are  more convenient for fitting:  
\beq
a = \frac{72\pi}{ \Lambda^2} \epsilon_{int} 
~~,~~
b \simeq \frac{\epsilon_{NP}}{ [\Lambda/(9.0 ~{\rm TeV})]^4}
~~,~~ c = \frac{1}{m^2}
%\simeq \left(\frac{29.6 \pi}{\Lambda^2}\right)^2 \epsilon_{NP} 
\label{eq2.1.14}
\eeq
We take $c,b$ positive, but $a$ can have either sign,
for destructive  or constructive interference with
the SM. 

The approximation of neglecting the
 ${\cal O}(g_2^2\lambda^2 \hats^2)$ term
(see the end of section \ref{sec:CIvsFF})
allows a clear distinction between 
the model-dependent constants
$\epsilon_{int}$, $\epsilon_{NP}$, and 
 the free parameters $\lambda,m$
which we wish to constrain.
In the next section, data  will  give 
allowed  ellipses  in $a,b$ space, for particular
values of  $c = 1/m^2$. A model 
(one remaining  parameter $\lambda$)
will correspond to a  parabola in $a,b$ space,
identified by the constants 
$\epsilon_{int}$ and $\epsilon_{NP}$
\beq
b \simeq   \frac{  \epsilon_{NP}}{8 \epsilon_{int}^2} a^2
%\frac{a}{\sqrt{b}  }\simeq\sqrt{\frac{32}{3K}}\frac{\epsilon_{int}}{\sqrt{\epsilon_{NP}}}\simeq 3.24 \frac{\epsilon_{int}}{\sqrt{K\epsilon_{NP}}}
\eeq
and the bound on $\lambda$ (or $\lambda^2/2m^2$) will
arise at the point where this parabola
leaves the allowed ellipse.

\section{Estimated bounds from  data?}
\label{sec:data?}

The usual way to set bounds on a model, is
to fix the parameters to representative values,
 simulate the expected signal, and
compare it to data.  Our $a,b,c$ parameters
could be constrained in this way, following
the recipe given in \cite{Jose}.
 Here,  we explore instead, naively 
to fit the difference between data and
SM expectation to a function, and
constrain that function.

CMS compared 20 fb$^{-1}$ of $pp \to \lplm$ data to the
NNLO SM predictions\cite{CMSMS}, and  in a separate
publication\cite{CMSLam}, obtained bounds
 on  the 
  CI of eq. (\ref{ATLAS}).
%for $\ell=e,\mu$ . 
We use the CI analysis,
and start with
the $\mpmm$ data  because it shows a slight preference for
NP with destructive interference  \footnote{The CMS study\cite{CMSLam}
does not appear to allow the CI
to reduce the number of events with respect to
the SM expectation, so is perhaps less sensitive to
this solution than our simplistic fit.}.
The  data is binned, which poses a problem  in comparing
to eq. (\ref{dream}), because the pdfs only  cancel in
the ratio at the same  value of $\hats$. We will 
quantify the resulting uncertainty in a later publication; 
here we take the uncertainly in $\hats$  to be the bin-width.
 
\begin{figure}[ht]
\unitlength.5mm
%\SetScale{1.418}
%\begin{boldmath}
\begin{center}
\epsfig{file=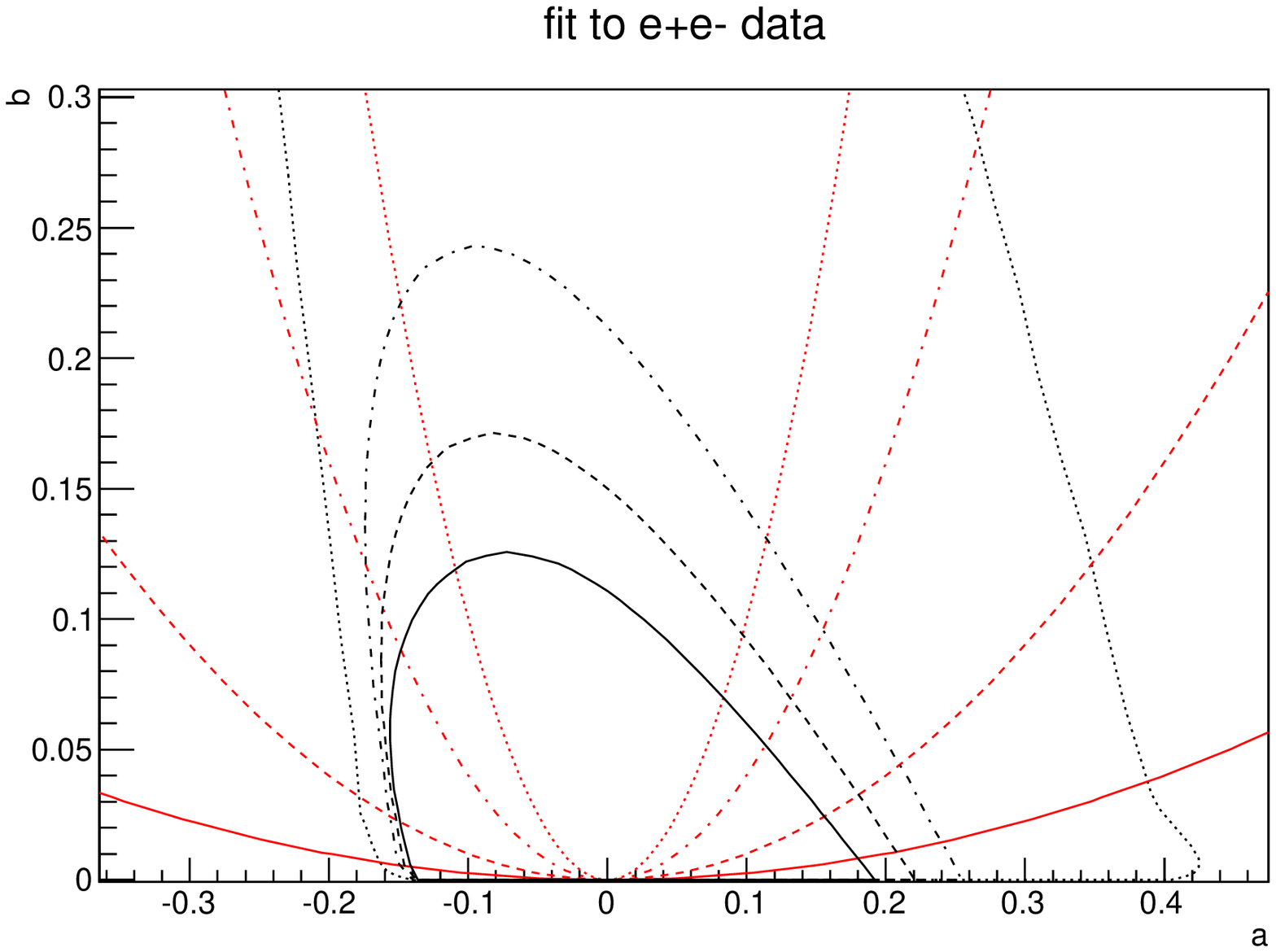,height=6cm,width=9cm}
\epsfig{file=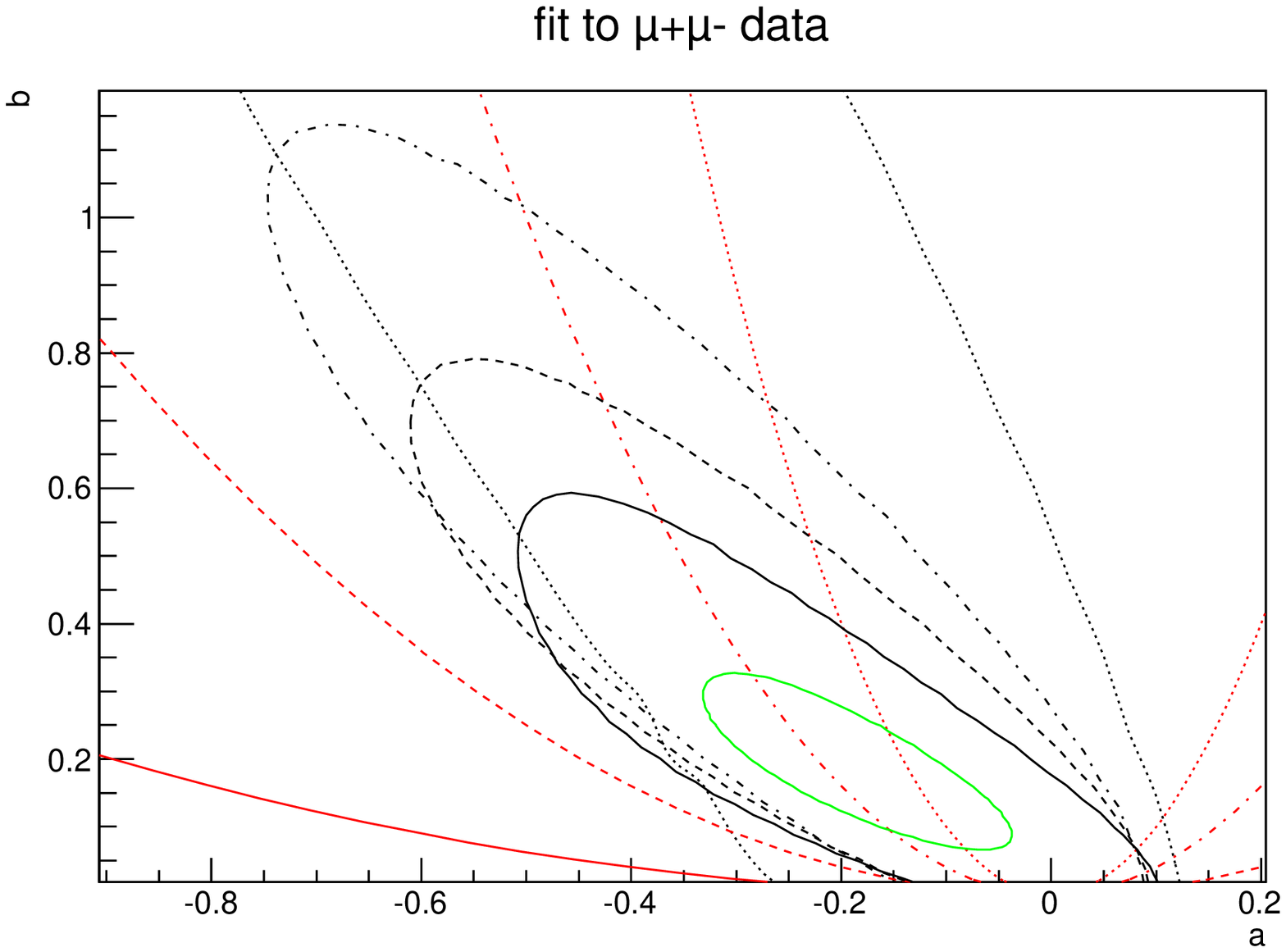,height=6cm,width=9cm}
\end{center}
%\end{boldmath}
\caption{ 
 $2\sigma$-allowed  ellipses for the $a,b$ parameters
of eq. (\ref{dream}), with fixed values of  $c= 1/m^2$, 
 obtained from $\epem$ (upper)
and $\mpmm$ (lower) data\cite{CMSLam}.
Decreasing ellipse size corresponds to  
$m = 1,2,3, \infty$ TeV.   The  dotted 1 TeV
 $\mpmm$ ellipse (lower plot),  extends to  $b\simeq 3.5$. 
The inner green ellipse (lower plot)
is the $1\sigma$-allowed contour for  the contact interaction. 
The parabolae, in decreasing
width,  represent models described by 
$b = 1/4a^2, a^2, 4 a^2, 10 a^2$.
% Fait avec ORDI/ROOT/FIT/dofitdeSmu3.C  
\label{fig:ell} }
\end{figure}

From 
the CMS plots, we read  the ratio  of data
to SM expectation
%\footnote{There are SM non-DY processes that produce $\lplm$, which
%are neglected when we identify this experimental
%ratio to $(d\sigma/\d\hats)/(d\sigma_{DY}/\d\hats)$.}
with  statistical and systematic uncertainties, 
for $\sqrt{\hats}> 300$ GeV.
We  perform a least-squares fit (minuit\cite{minuit})
of  $a$ and $b$  %of  eq. (\ref{dream}),
to these points
 for four  fixed values of $c= 1/m^2$  corresponding
to $m = 1,2,3$ TeV and $m\to \infty$. The resulting $2\sigma$
ellipses are plotted in figure \ref{fig:ell}.  They lean 
left because the destructive interference  of a
negative $a$-term 
partially cancels the excess  events generated by a
positive $b$-term.

 To extract bounds on a particular
model from this plot, recall that the form factor
parametrisation allows  to constrain  both
the coupling and mass  of an exchanged particle,
rather than the CI combination
$\sim \lambda^2/m^2$. However, in the interests
of performing a linear fit, we fixed the
value of the mass and  fit
to $a$ and $b$. 
The recipe to extract a bound is therefore
to   choose an ellipse labelled by a mass,
then   identify
the parabola 
which represents the model.
 The value of $b$ where the parabola
leaves the ellipse can be combined
with eq. (\ref{eq2.1.14}) to constrain
the CI induced
in the model, which  for finite $m$, 
gives a bound on $\lambda$.

The CI of eq. (\ref{ATLAS})
corresponds to the dash-dotted parabola, and the
inner ellipse, so the above recipe,
applied to the $\mpmm$ plot, gives 
$ \Lambda_{des} \gsim  10.3$  TeV and 
$\Lambda_{cons} \gsim $ 20.7  TeV, 
to be compared with the  CMS 
95\% C.L. exclusions:
$ \Lambda_{des} \gsim  12.2$  TeV,
$\Lambda_{cons} \gsim $ 15.0  TeV.
The difference  is partially due to
the data's  slight deviation from the SM:
%corresponds to the  model  of eq.(\ref{ATLAS}), and 
the dott-dashed  parabola of 
 eq. (\ref{ATLAS})
passes close to the central value
of the CI fit,
%$a,b \simeq -0.18,0.18$, 
which  translates   to $\Lambda_{des} = 13.7$ TeV.
If instead we estimate an expected limit
(by centering the ellipses  on the origin),
we obtain  $\Lambda_{des} \geq 12.6  $ TeV,
 $\Lambda_{cons} \geq 18.1 $ TeV, to be
compared to CMS's expected limits of 13 and 17 TeV. 
In the case of the $\epem$ data,  our
estimates give  $\Lambda_{des} \geq 16.3  $ TeV,
 $\Lambda_{cons} \geq 19.0 $ TeV, to be
compared to CMS's bounds of  13.5 and 18.3 TeV.

Finally, we can estimate bounds on  the
first generation leptoquark 
 $S_0$, interacting with $\overline{u^c} e$. 
This model corresponds roughly to the dashed  parabola,
and bounds can be obtained from ellipses
corresponding to the $\epem$ data (see figure
\ref{fig:ell}). The current
LHC direct production  bound on the mass of leptoquarks
decaying to electrons  is 830 GeV \cite{CMSLQ},
so it is interesting to study the
couplings of 
leptoquarks with  masses of 1,2 and 3 TeV.
We obtain 
 $\Lambda_{des} \geq 21.0  $ TeV,
 $\Lambda_{cons} \geq 20.3   $ TeV in the contact interaction limit, and
$|\lambda_R| \lsim 0.3, 0.5 $ and 0.75 for $m= 1,2,3$ TeV. 
Retaining the
${\cal O}(g^2 \lambda^2 \hats^2)$
term of eq. (\ref{LQff}) gives
$\lambda_R \lsim 0.4$ at $m=1$ TeV. These bounds are
stronger, by  factors of  few,
than  those obtained  in \cite{WZ} by
simulating leptoquark exchange in MadGraph,
and excluding parameters that  generate excess events. 
We imagine that our more restrictive bound could arise
from requiring that the leptoquark's destructive interference
not reduce the event rate at $\sqrt{\hats} \lsim$ TeV.

\section{Summary}

The differential cross-section $d\sigma/d\hats (pp \to \lplm)$ can almost
be written as a partonic cross-section  multiplied by an integral
over parton distribution functions (see section \ref{sec:ptoFF}).
When the partonic cross-section includes the  exchange of a heavy
new particle in the $t$-channel, such as a leptoquark, then,
as shown in section \ref{sec:CIvsFF},
it is better approximated by  the ``form factor'' 
 expansion in $\hats/(\hats + m^2)$,
than by  the usual local expansion in $\hats/m^2$ which gives
contact interactions.
 If experimental limits were set on
the coefficients of  the  form factor parametrisation given
in eq. (\ref{dream}), these could be translated to 
any combination of contact interactions, or the
exchange of an arbitrary leptoquark, by a simple
analytic calculation.

\begin{acknowledgments}

We thank Gian Giudice,  Maxime Gouzevitch, and  Michelangelo
Mangano  for
useful discussions, and Robbie Burns and J. Steinbeck
for the title.
\end{acknowledgments}

\end{document}